\documentclass[11pt,twoside]{article}
\usepackage{asp2010}

\resetcounters 

\begin{document}

\markboth{Cross et al.}{Automated curation of the VSA}

\title{Automated curation of infra-red imaging data in the WFCAM and VISTA
Science Archives} 
\author{Nicholas~Cross$^1$, Ross~Collins$^1$, Eckhard~Sutorius$^1$,
Nigel~Hambly$^1$, Rob Blake$^1$, and Mike Read$^1$.
\affil{$^1$Scottish Universities' Physics Alliance (SUPA),Institute for Astronomy, 
School of Physics, University of Edinburgh,
Royal Observatory, Blackford Hill, Edinburgh EH9 3HJ, UK}}

\begin{abstract}
The two fastest near infrared survey telescopes are UKIRT-WFCAM and VISTA. The 
data from both these instruments are being archived by Wide Field Astronomy 
Unit (WFAU) at the IfA, Edinburgh, using the same curation pipeline, with some
instrument specific processing. The final catalogues from these surveys will 
contain many tens of billions of detections.

Data are taken for a range of large surveys and smaller PI programmes. The
surveys vary from shallow hemisphere surveys to ultra deep single pointings
with hundreds of individual epochs, each with a wide range of scientific goals,
leading to a wide range of products and database tables being created.
Processing of the main surveys must allow for the inclusion of specific
high-level requirements from the survey teams, but automation reduces the amount
of work by archive operators allowing a higher curation efficiency. The decision
making processes which drive the curation pipeline are a crucial element for 
efficient archiving. This paper describes the main issues involved in automating the
pipeline.
\end{abstract} 
 
\vspace{-10mm} 
\section{The WFCAM and VISTA Science Archives}
The WFCAM and VISTA Science Archives \citep{Hamb08} are the main access for
data from WFCAM \citep{Cas07} and VISTA \citep{Emer10}. The majority of
time on both instruments is spent on large surveys: UKIDSS \citep{Lawr07}, and
the VISTA Public Surveys \citep{Arna07}. There are also a range of smaller
Principal Investigator (PI) programmes allocated by the Telescope Allocation 
Committees each semester that require curating. We run the same set
of tasks on all the surveys and programmes, with the amount of processing in each 
task dependent on the type of programme. For instance a wide survey will spend
more time on band-merging and neighbour tables, but a deep survey will spend more time on 
deep stack creation and multi-epoch tables. PI programmes are set up completely
automatically\footnote{Occasionally we have manually grouped together several
related PI programmes from different semesters before automatic processing.}
because of the large number of programmes, but surveys are set up in a
semi-automatic way receiving special instructions, quality control and sometimes 
additional products from the science teams. PI programmes usually obtain all 
their data in one observing semester and are processed completely at the end of
the semester, so new releases will be due to software or calibration
improvements necessitating a complete reprocessing. Surveys build up data
over many semesters and will be appended to, as well as occasionally
reprocessed.  These different scenarios have to be factored into the pipeline control.

\section{Overview of data pipeline}
\label{overview}
Data consisting of images and catalogues are first transferred
to WFAU by the Cambridge Astronomy Survey Unit (CASU) who process each observing
block and calibrate the data. We ingest these into the science archives along
with any external data or quality control provided by the science teams for
public surveys. The automated curation pipeline is then run, executing the following tasks:
\begin{itemize}
  \item Quality Control
  \item Programme setup (using {\bf ProgrammeBuilder} class)
  \item Creation and ingestion of deep products and catalogues
  \item Creation of band-merged Source table from deepest products
  \item Recalibration of each epoch 
  \item Creation of band-merged catalogues for each epoch
  \item Creation of neighbour tables 
  \item Creation of synoptic tables for light-curves and variability analysis
\end{itemize}

The dataflow plan for automated curation of a single programme is shown in
Fig~\ref{fig:dataFlows}a. The automated pipeline has changed in a couple of
ways since \cite{Coll09}. We have removed the distinction between deep and
shallow programmes in a way that also allows us to do an easier comparison
between the curation and data tables and removes the need to copy data.
Appending either in width (more pointings) or depth (more epochs) is much
easier. We have also designed a much more sophisticated SQL schema template 
that can be used by both surveys and PI programmes so that only one template is
needed per instrument. These changes have significantly increased the amount of
automation of public surveys so that many of the curation tables are filled in 
the same way as PI programmes.
 
\section{Setting up a programme}
\begin{figure}
\includegraphics[clip,width=60mm,angle=0]{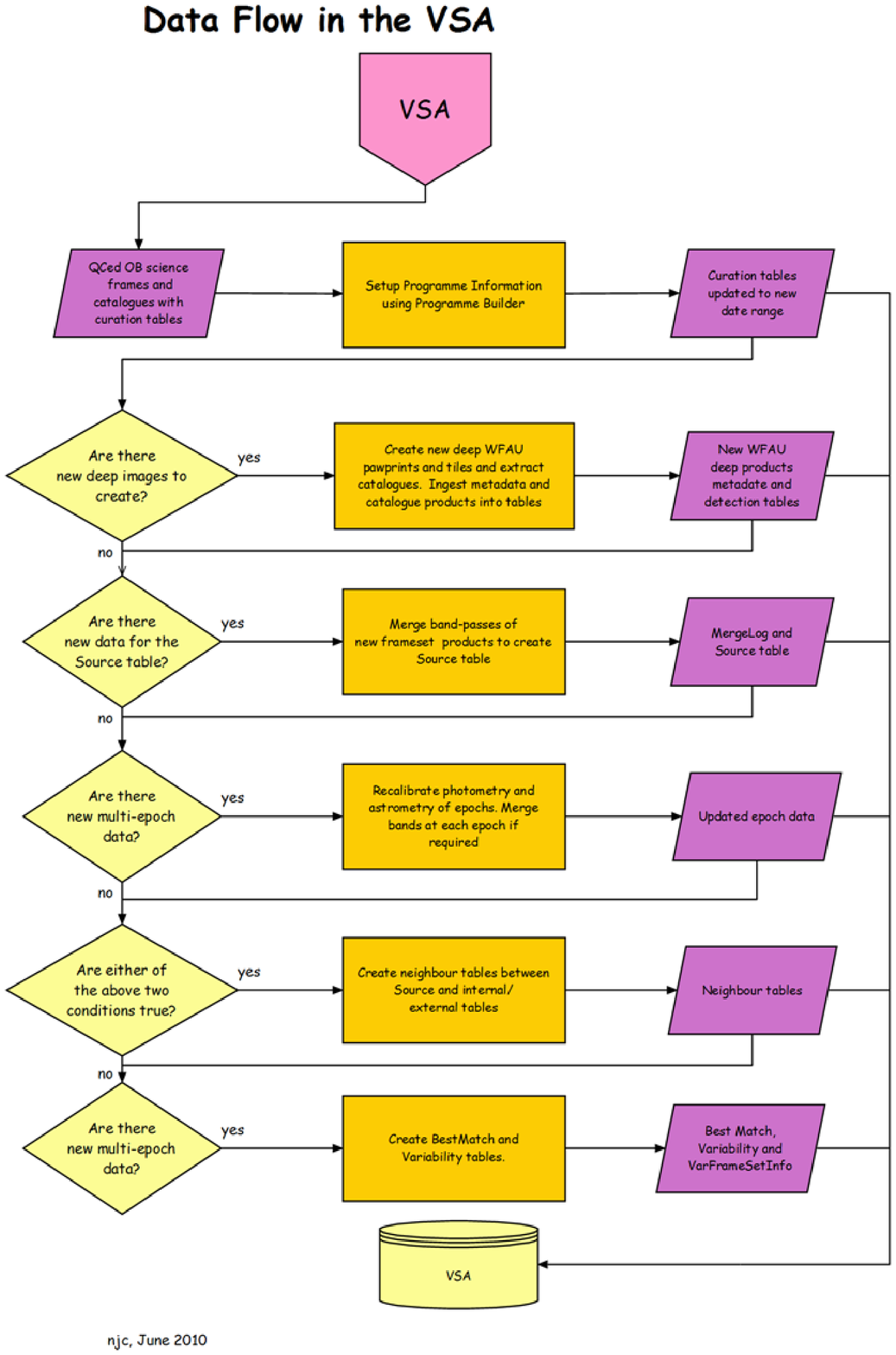}
\includegraphics[clip,width=60mm,angle=0]{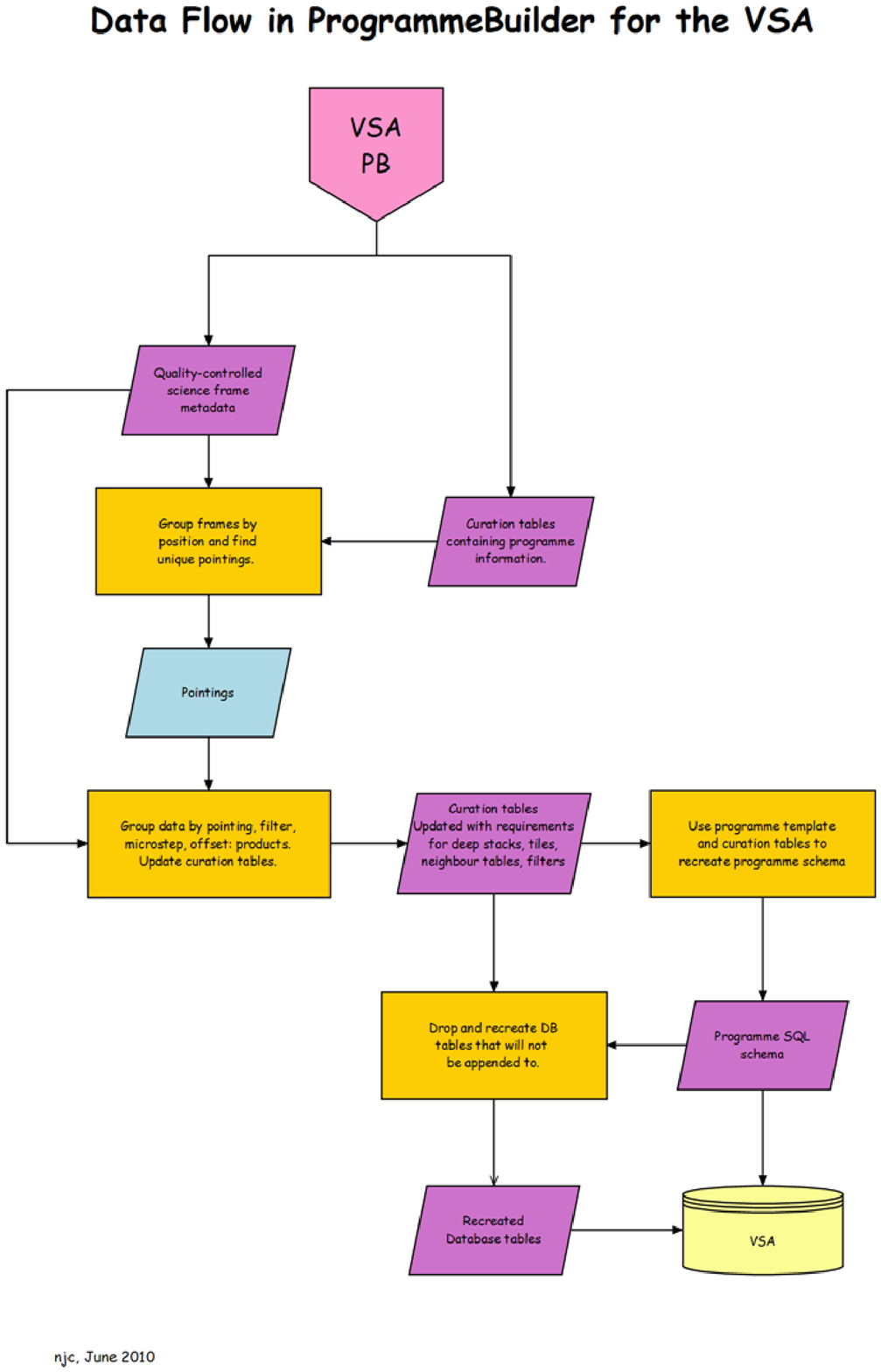}
  \caption{Left (a): Overall dataflow for WSA or VSA automated processing.
  Right (b): Dataflow for the ProgrammeBuilder class. In both figures, the
  rectangular yellow boxes represent tasks, the parallelograms represent data
  (light blue for temporary and purple for permanent) and the light yellow 
  diamonds represent control structures. The cylinder represents the database 
  that the permanent data are ingested into and the pentagon represents the 
  pipeline that the dataflow is embedded in. In Fig a), processing of each task
  to create a permanent data object is determined by a control structure. New
  data feeds into the control for the next task. Each control structure compares
  expected products, based on the sorting and grouping of data in Fig b), to
  actual created products and before determining what still needs to be
  created.}
  \label{fig:dataFlows}
  \vspace{-5mm}
\end{figure}

The curation tables that control the pipeline are filled by the 
{\bf ProgrammeBuilder} class, see Fig~\ref{fig:dataFlows}b. The image metadata
is used to group pawprint frames by position, filter, microstepping and in the 
case of VISTA position angle and offset position within a standard tile. Unique
pointings are found by grouping by position alone and then products are found
for each filter in these pointings. The information for each pawprint product 
is put into the \verb+RequiredStack+ table. For VISTA, the pawprint stacks are
grouped into tiles in \verb+RequiredTile+ and the two tables are linked via 
\verb+ProductLinks+. For each filter used, the number of epochs at each pointing
is found, and this is used to determine whether multi-epoch tables are required.
This information goes into \verb+RequiredFilters+, \verb+Programme+ and 
\verb+RequiredNeighbours+.

\newpage

\verb+RequiredNeighbours+ contains the list of all
the neighbour and cross-match tables which need to be created. In the case of PI
programmes, a common set of internal neighbour tables and cross-matches to
all-sky surveys is produced and in addition surveys are cross-matched with wide
range of specified external surveys.
 
Once these curation tables are setup, a programme specific schema is created
using them and a template SQL schema. The template schema is composed of SQL
definitions for different attributes, substitution strings and control structures. 
We give an example below from the template for VISTA \verb+Source+ tables:
 
\begin{verbatim}  
++c:a
**s*&a&m&b&Pnt      real not null,      --/D Point source colour 
&As&-&Bs& (using aperMag3)  --/U mag  --/C PHOT_COLOR  --/Q 
&a&AperMag3,&b&AperMag3  --/N -0.9999995e9  --/G 
allSource::colours
==c:a
\end{verbatim}

The ${\rm ++c:a}$ line is a control structure which repeats each subsequent line
for all the programme filters until the ${\rm ==c:a}$ line for all colour
combinations in the survey. The ${\rm **s*}$ structure controls which table(s)
each line goes into when several narrow tables are created for curation purposes
and subsequently joined at release into one table. Lines will only go into the 
\verb+Source+ table, but not the \verb+MergeSource+. ${\rm \&a\&}$, ${\rm
\&A\&}$, ${\rm \&b\&}$ and ${\rm \&B\&}$ are substitution strings, where ${\rm
a}$ and ${\rm b}$ refer to the first and second filter in a colour respectively
and ${\rm a}$ and ${\rm A}$ refer to lower and upper case respectively. When the
template is processed for the VISTA-VMC (Cioni, M.-R. et al. 2011, in
prep.),which contains Y, J and Ks band data, the following piece of schema is
produced:

\begin{verbatim}
ymjPnt      real not null,      --/D Point source colour Y-J 
(using aperMag3)  --/U mag  --/C PHOT_COLOR  --/Q yAperMag3,
jAperMag3  --/N -0.9999995e9  --/G allSource::colours
jmksPnt      real not null,      --/D Point source colour J-Ks 
(using aperMag3) --/U mag  --/C PHOT_COLOR  --/Q jAperMag3,
ksAperMag3  --/N -0.9999995e9  --/G allSource::colours 
\end{verbatim}

The SQL schema is used to create the database with the correct tables, control
the code that produces the table data and create a schema browser and glossary
for scientists. Having a single template and control structures reduces the
need to repeat the SQL, which makes it much easier to update and maintain.

\section{Triggering the automated pipeline}

Each task is triggered by comparing the curation tables with the data tables.
For instance, the deep product curation task will compare the products specified in \verb+RequiredStack+ and
\verb+RequiredTile+ with data in the tables \verb+Multiframe+ and
\verb+ProgrammeFrame+ matching on the programme, the product identifier and the
release number. If all required products have been created then the pipeline
moves onto the next task. If not, the remaining products will be created. Thus
the pipeline can be restarted easily if there is a network error or software
bug. Other curation tasks are similarly triggered and the data tables are
updated at the end of each task. Log files are produced and curation history
tables are kept up to date, so any failures can easily be identified.

\bibliographystyle{asp2010}
\bibliography{P42}

\end{document}